\pdfoutput=1
\documentclass[sigconf, authorversion, 10pt]{acmart}
\usepackage{booktabs}
\usepackage{url}
\makeatletter
\g@addto@macro{\UrlBreaks}{\UrlOrds}
\makeatother
\usepackage[utf8]{inputenc}
\usepackage{prettyref}
\newrefformat{sec}{Section~\enquote{\nameref{#1}}}
\newrefformat{fig}{Fig.~\ref{#1}}
\newrefformat{tab}{Table~\ref{#1}}
\graphicspath{{images/}}
\usepackage{array}
\usepackage{multirow}
\newcolumntype{L}[1]{>{\raggedright\arraybackslash}p{#1}}
\newcolumntype{C}[1]{>{\centering\arraybackslash}p{#1}}

\def \etal {et al}
\usepackage{url}
\usepackage{color, colortbl}
\colorlet{tabcolor}{gray!15}
\usepackage[caption=false]{subfig}
\usepackage{relsize}
\usepackage{amsmath}
\usepackage{commath}
\usepackage{textcomp}
\usepackage{xspace}
\usepackage{siunitx}

\usepackage{multirow}

\newcommand{\ssymbol}[1]{^{\@fnsymbol{#1}}}

\setcopyright{none}

\renewcommand\footnotetextcopyrightpermission[1]{}

\usepackage{textcomp, libertine}
\acmDOI{10.1145/3278161.3278166}

\begin{document}
	\title{5G Applications: Requirements, Challenges, and Outlook}
	
	\author{Aaron Yi Ding}
	\affiliation{%
		\institution{Delft University of Technology, Netherlands}
	}
	\email{aaron.ding@tudelft.nl}
	
	\author{Marijn Janssen}
	\affiliation{%
		\institution{Delft University of Technology, Netherlands}
	}
	\email{m.f.w.h.a.janssen@tudelft.nl}
	
	\renewcommand{\shorttitle}{5G Applications}
	\renewcommand{\shortauthors}{Aaron Yi Ding, Marijn Janssen}
	
	\begin{abstract}
		%!TEX root = ../main.tex

The increasing demand for mobile network capacity driven by Internet of Things (IoT) applications results in the need for understanding better the potential and limitations of 5G networks. Vertical application areas like smart mobility, energy networks, industrial IoT applications, and AR/VR enhanced services all pose different requirements on the use of 5G networks. Some applications need low latency, whereas others need high bandwidth or security support. The goal of this paper is to identify the requirements and to understand the limitations for 5G driven applications. We review application areas and list the typical challenges and requirements posed on 5G networks. A main challenge will be to develop a network architecture being able to dynamically adapt to fluctuating traffic patterns and accommodating various technologies such as edge computing, blockchain based distributed ledger, software defined networking, and virtualization. To inspire future research, we reveal open problems and highlight the need for piloting with 5G applications, with tangible steps, to understand the configuration of 5G networks and the use of applications across multiple vertical industries.
	\end{abstract}


\maketitle

\section{Introduction}
\label{sec:introduction}
The success of mobile communication stems from its pervasive coverage and substantial ecosystem that boost rapid pace of innovation in terms of new applications and venture creations. To withstand its long-term prosperity, the upcoming fifth generation (5G) of mobile networks are expected to generate new opportunities in the era of Internet of Things (IoT), autonomous driving, augmented and virtual reality (AR/VR) services. This vision is supported by the on-going development of 5G cellular architecture and its air interface enhancement \cite{Wang.2014, Akpakwu.2018} to cater for massive deployment of connected devices, which are projected to reach more than 75 billion by 2025 \cite{Statista.2017}.

\begin{figure*}
	\centering
	\includegraphics[width=0.9\linewidth]{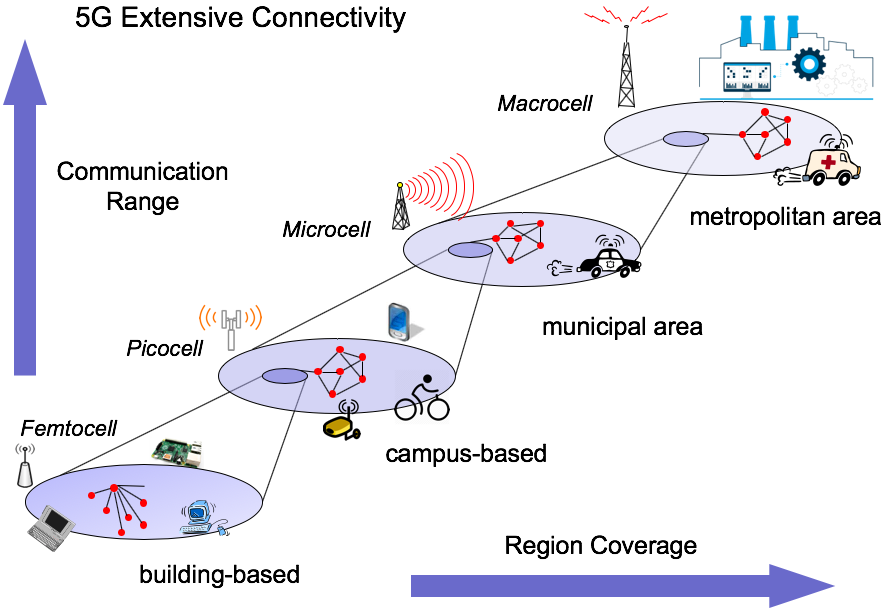}
	\caption{Envision of 5G Network Connectivity}
	\label{fig:overview}
\end{figure*}

Comparing with existing 4G, 5G networks encompass new wireless interfaces to support higher frequencies and spectrum efficiency. There is significant improvement in terms of signaling, management and accounting procedures at the 5G core networks in order to accommodate the needs from diverse range of new applications that are outside traditional mobile broadband category \cite{ngmn.2015}. By its design, 5G deployment will provide extensive connectivity through its heterogeneous wireless access, ranging from macrocell (long range) to femtocell (short range). As shown in Figure \ref{fig:overview}, the coverage will span across metropolitan area, municipal area and down to campuses and buildings. This pervasive connectivity is the key to seamless mobility and service availability that has been centered in the cellular system since its debut. 

Given the new demands from IoT, autonomous driving, AR/VR and smart city services, one important pursuit for 5G is to match its capacity to the scale and growth of various 5G driven applications in an economical and sustainable manner. This mission covers network architecture, communication techniques, ecosystem design and actual deployment. Recent efforts have sought the utilization of network function virtualization (NFV), software-defined networking (SDN), edge computing and offloading, as well as distributed data analytics (e.g., Apache Spark \cite{Spark.2018}). Those technical innovations have shown promising results \cite{Han.2015, Ding.2014, Costa.2014, Ding.2015a, Ding.2015, Ding.2015b, Flores.2017, Morabito.2018, Cozzolino.2017, Cozzolino.2018, Pu.2015}. Meanwhile, as new application domains are inspired by 5G on almost daily basis, we still lack a comprehensive understanding on requirements originating from various domains which include both technical and governance perspectives, especially what opportunities and challenges 5G will endure in the current transitional phase. 

Motivated by the latest advance in 5G and trend of urbanization \cite{UnitedNations.2014}, this paper tackles the challenges of 5G from the application perspective. In particular, we focus on vertical application domains, which are built for target enterprise and entities with specific requirements. As a solid step to demystify the requirements in the context of IoT and smart cities, we aim to answer the major question: 
\textbf{\textit{"How do we consolidate 5G driven applications across multiple vertical industries to unveil the full potential of 5G?"}}
Besides identifying challenges, our work also highlights the opportunities for 5G applications in various vertical domains, which can shed light on future development for researchers, engineers and policy makers from academia, industry and government.

Our key contributions are hence twofold:

\begin{itemize}
	\item First, we classify the application domains inspired by 5G, and quantify their key requirements. Our requirement analysis covers major aspects including communication range, bandwidth capacity, latency, reliability, energy, security and privacy.
	\item Second, we pinpoint open challenges and opportunities based on reviewing the state-of-the-art research and project initiatives. Our discussions cover both technical aspect and also regulation and governance. We further stress the need to pilot 5G experimental testbed through tight cooperation across universities, network operators, equipment vendors and governmental institutes. 
\end{itemize}

We note that although this work provides an extensive sampling of existing and emerging 5G applications, our study does not attempt to cover every nuance. The rest of this paper is organized as follows. Section 2 provides an overview of application domains in 5G. Section 3 illustrates the application requirements. Section 4 highlights open challenges and potential opportunities in 5G. We discuss related work and project initiatives and conclude with our outlook in Section 5.

\section{5G Enabled Applications}
\label{sec:application}
The advances in mobile networking have created a myriad of diverse applications to improve the life quality of end users, including smart mobility, digital commerce, social networking and health care. From a broader perspective, mobile applications are part of the Internet services, which have witness a rapid evolution over the past decades.

\begin{figure*}
	\centering
	\includegraphics[width=\linewidth]{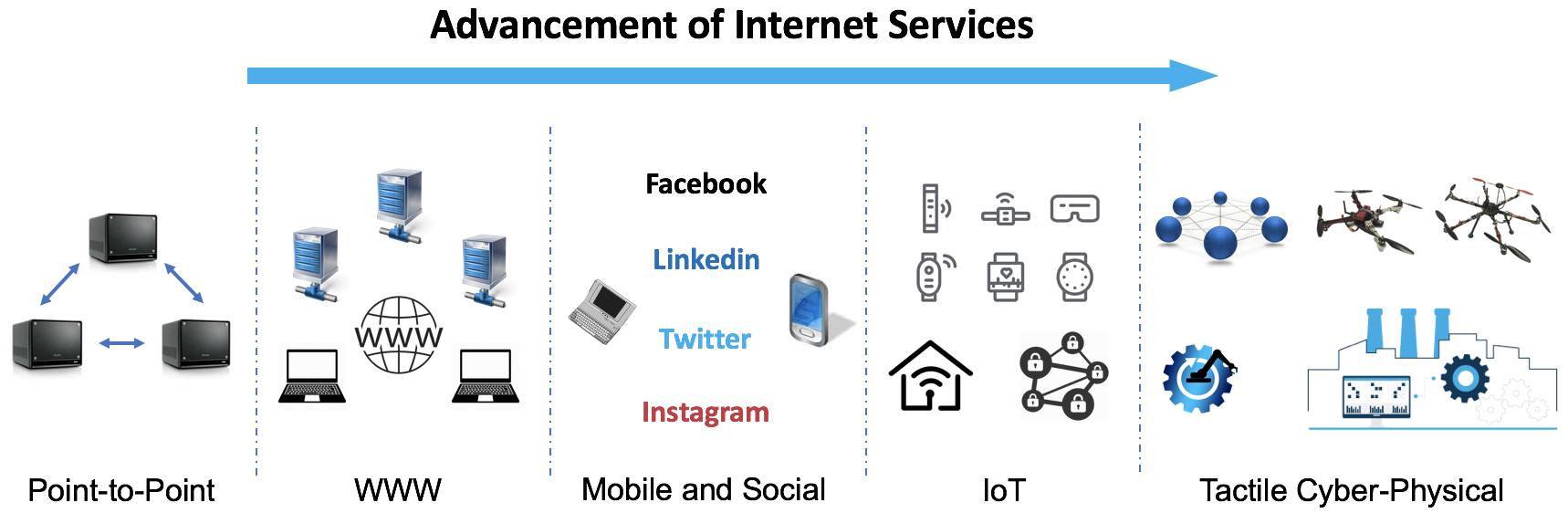}
	\caption{Advancement of Internet Services}
	\label{fig:services}
\end{figure*}

As illustrated in Figure \ref{fig:services}, the Internet services have evolved from conventional point-to-point data exchange, world wide web (WWW), mobile and social applications, to the recent IoT services and forthcoming tactile Internet \cite{Zanella.2014, Fettweis.2014, Simsek.2016}. In specific, the tactile Internet applications will facilitate the integration between digital sphere and our physical environments, covering advanced use cases of machine-to-machine (M2M) communication. Those new applications are characterized by the need for a network having ultra low latency, high availability, reliability and security. Many of these applications are also context-aware where the context is sensed for triggering actions, e.g., smartphones are nowadays aware if the owner is driving and can avoid interrupting the driver \cite{Shishkov.2018}. 

Among several application domains, the IoT-empowered smart city has become a focal concern of 5G. In this context, smart city integrates traditional and modern Information and Communication Technology (ICT) for a unified and simple access to services for the city administration and the residents. The aim is an enhanced use of resources, improving quality of services for citizens while reducing operational costs of public administration \cite{Anthopoulos.2015} and reducing the administrative burden for citizens and businesses. For example, smart transportation should reduce congestion and pollution and at the same time result in higher utilization of transport.

On the one hand, IoT has quickly advanced from an experimental technology to the driving force of 5G systems. To fully exploit the opportunities behind IoT, 5G has placed IoT at a vital position in its ecosystem. On the other hand, realizing the IoT vision of smart city depends on a careful integration with 5G telecommunication technologies to provide scalable and robust connectivity. Comprehensive and scalable supports from 5G are hence required to overcome the economic and technical constraints of state-of-the-art conceptualizations and implementations, while maintaining both practical and commercial appeals.

%
%\begin{figure}
%	\centering
%	\includegraphics[width=0.95\linewidth]{fig/integration}
%	\caption{Integration of 5G and Service Infrastructures}
%	\label{fig:integration}
%\end{figure}

%As highlighted in Figure \ref{fig:integration}, 
For 5G driven applications, we highlight five domains that can benefit from a tight integration with 5G and next generation cyber-physical infrastructure.

\begin{figure*}
	\centering
	\includegraphics[width=\linewidth]{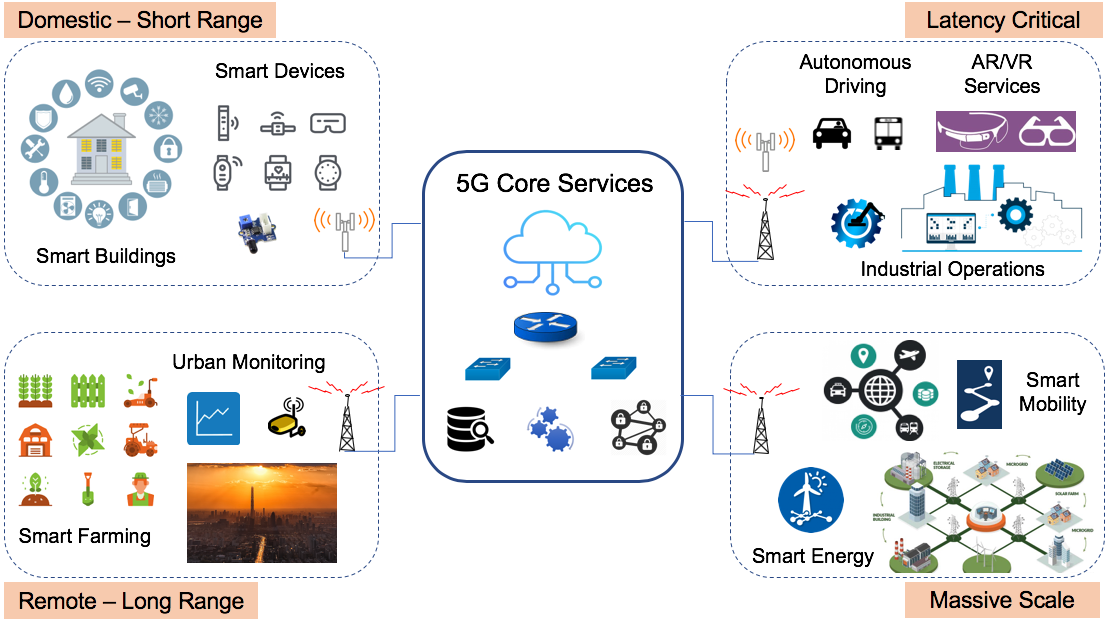}
	\caption{Application groups empowered by 5G}
	\label{fig:5g-app}
\end{figure*}

\begin{itemize}
 
  \item \textbf{Smart Mobility}: Mobility applications in 5G ranges from traditional road/route planning to the emerging autonomous driving services (connected vehicles) and extended sharing economics of smart transportation. The benefits of smart mobility include traffic balancing, efficient routing, accident prevention, energy saving, cost and emission reduction \cite{Benevolo.2016}. From this group of applications, there is a strong demand for 5G to support pervasive connectivity, low latency, high speed and link reliability, security and low power consumption.
 
 \item \textbf{Smart Energy}: This category of applications includes power plant monitoring and management, smart grid networking, power failure detection and response, new consumption saving services for homes and office buildings, energy marketplace and smart charging stations for electronic vehicles. Smart energy is expected to enhance efficiency and reliability of power systems with renewable energy and achieve intelligent distribution. The major demands for 5G are on link reliability, security and privacy \cite{Fang.2012, Wang.2013}.

 \item \textbf{Smart Health}: Health applications are becoming popular among mobile users owing to the growing awareness of fitness and well-being. Together with the advance of smart wearables, applications of this category have covered mobile based condition monitoring and diagnosis, environmental quality inspection. With more data collected from sensors deployed on wearable devices, smart health will positively influence the medical and healthcare systems \cite{Akpakwu.2018}. Another emerging application in this domain is the AR/VR enabled surgery, which will demand low latency and high bandwidth, on top of the general requirements of low power, security and data privacy from 5G. 
 
 \item \textbf{Industrial Applications}: Applications such as Industry IoT 4.0 \cite{Wang:2016} represent the next generation of cyber-physical services in terms of manufacturing, machine-to-machine (M2M) communication, 3D printing and AI supported construction. The impact of those industrial applications will extend beyond factories and plants, directly benefiting the entire society. The major demands for 5G include critically high reliability, ultra low latency, support of massive deployment, security and privacy.

 \item \textbf{Consumer Applications}: The vast amount of consumer applications (Apps) reflect the potential of 5G mobile business and technology innovations. As we are familiar with typical mobile applications running on smartphones and tablets, the emerging applications include ultra HD (4K/8K) mobile streaming, blockchain based financial technology (FinTech), pervasive gaming (like Pokemon GO \footnote{https://www.pokemongo.com/en-us/}), mobile AR/VR mixed reality services supported by unmanned aerial vehicles, and holographic technology such as HoloLens \footnote{https://www.microsoft.com/en-us/hololens}. All those advanced services are demanding 5G to support extensive connectivity, high bandwidth, low latency, low energy footprint, link reliability and security.

\end{itemize}

\section{Requirement Analysis}
\label{sec:evaluation}
To carry out a fine-grained analysis that reflects technical interdependency, we break down the aforementioned application domains (detailed Section 2) into four distinct types. As shown in Figure \ref{fig:5g-app}, 5G driven applications are divided into four categories: 1) domestic type with short communication range, 2) remote type with long range, 3) latency critical, and 4) massive scale. Our goal is to quantitatively manifest their requirements for facilitating future development and deployment of 5G systems. We also note that although this grouping includes a wide range of existing and emerging applications, our discussions do not attempt to cover every nuance.

\begin{table*}[h]
	\centering
	\caption{Vertical Application Requirements in 5G}
	\label{tab:requirement}
	\begin{tabular}{|p{3cm}|L{2.5cm}|L{2.6cm}|L{2cm}|L{1.8cm}|L{1.2cm}|L{1.7cm}|}
		\hline
		{\textbf{\centering{Applications}}}
		&\textbf{{Communication Range}}
		&\textbf{{Bandwidth Capacity}}
		&\textbf{{Latency}}
		&\textbf{{Link Reliability}}
		&\textbf{{Energy}} 
		&\textbf{{Security Privacy}}\\ \hline \hline

		\rowcolor{tabcolor}
		
		Smart buildings   & short range  &  10 - 1000~Mbps   &  Median & Median & Low & High \\
		\hline
		\rowcolor{tabcolor}
		Smart devices   & short range  &  10 - 1000~Mbps &  Median & Median & Low & High \\
		\hline

		Smart farming   & long range  &  1 - 100~Mbps  &  Tolerant & Median & Low & Median \\
		\hline
		
		Urban monitoring   & long range  &  1 - 100~Mbps  &  Tolerant & Median & Low & Median \\
		\hline
		
		\rowcolor{tabcolor}
		
		Autonomous driving   & long range  &  10 - 5000~Mbps &  Critical & High & High & Critical \\
		\hline
		\rowcolor{tabcolor}
		AR/VR services   & short range  &  100 - 5000~Mbps   &  Critical & Median & High & Median \\
		\hline
		
		Smart energy   & median range  & 10 - 1000~Mbps  &  Median & High & Median & High \\
		\hline
		
		Smart mobility  & long range  &  10 - 1000~Mbps  &  Median & High & Median & Median \\
		\hline
		
	\end{tabular}
\end{table*}

\subsection{General Requirements}

For each type of applications, we highlight the general requirements in Table \ref{tab:requirement}, covering communication range, bandwidth capacity, latency, link reliability, energy consumption, security and privacy.

\subsubsection{Domestic - short range}

As shown in Figure \ref{fig:5g-app}, this group of applications include consumer applications in the context of smart homes and office buildings \cite{Anthopoulos.2015}. Owing to their communication pattern, 5G needs to support low power networking, which is crucial for wearable devices. Given the exposed security issues at smart homes \cite{Hafeez.2016}, there is a strong demand to regulate unwanted traffic on the wireless interfaces.

\subsubsection{Remote - long range}

Applications in smart farming and urban monitoring demand 5G support especially in terms of communication coverage. Since devices deployed for farming and urban monitoring need to operate over long time period, energy saving is another key requirement.

\subsubsection{Latency critical}

Industrial applications are typically tied to safety in manufacturing and hence demanding high level of security. For consumer domain such as autonomous driving and AR/VR services, low latency (critical level) and high bandwidth must be supported in 5G communication. Due to the safety concern, autonomous driving also demands high link reliability.

\subsubsection{Massive scale}

For scenarios of massive deployment such as in smart grid and transportation systems, 
5G needs to elastically scale, to cater for increased traffic demand, number of end devices, and applications, and with acceptable cost. In particular to smart energy, high link reliability and security are also required.

\subsection{Requirements from Emerging Services}

For emerging applications in both smart city and vertical industries, 5G architecture needs to consider the requirements from several new angles. The first one rises from the swift of traffic pattern from downlink driven to uplink driven. This is mainly due to the introduction of high volume of data generated from smart vehicles, drones, and industrial IoT deployment. The traffic patterns might change and demand from different vertical industries and can shift over time. Being able to adapt to the fluctuation will be a key requirement for the 5G networks.

Secondly, due to the enforcement of General Data Protection Regulation (GDPR) \footnote{https://www.eugdpr.org/}, data privacy is becoming an avid issue. Data referring to an identifiable or identified person falls under this regulation. Especially with more embedded devices and autonomously flying drones/robots to collect data for surveillance purposes, 5G needs to guarantee security in communication and ensure privacy-by-design. The latter refers to ensuring data protection by having a proper architecture. 

Besides technical requirements, 5G must take into account the requirements from governmental and economical angles. In this context, connectivity of 5G in the future will be regarded as one of the mandatory common-pool resources (CPR) similar to water and electricity. This has strong implication on the regulation and management of 5G networks in terms of interoperability across operators, cost of maintenance, public-private sector ownership, wireless spectrum bidding and allocation (especially above 3 Ghz). Being a public resource, safety of large-scale operations will also become a key requirement.

%\subsection{The need for adaptation}
%In summary, the requirements differ per vertical application area and might change over time. One vertical industry might be demanding at a certain time, whereas another might be less demanding. The specific requirement might become more stringent over time. This all requires an architecture able to deal with fluctuations and changes in requirements. For this all kind of new technologies need to be incorporated in 5G networks.  

%\begin{itemize}
%
% \item Tactile Internet
% \item Sharing Economics with Autonomous Vehicles
% \item Commercial and Industrial Usage of Drones, Autonomously flying drones/robots with potential uses in crop pollination, search and rescue missions, surveillance, as well as high-resolution weather, climate and environmental monitoring
% \item Mobile 4K/8K AR/VR
%
%\end{itemize}

\section{Opportunities and Open Challenges}
\label{sec:discussion}
\subsection{Technology Opportunities}

\begin{figure*}
	\centering
	\includegraphics[width=\linewidth]{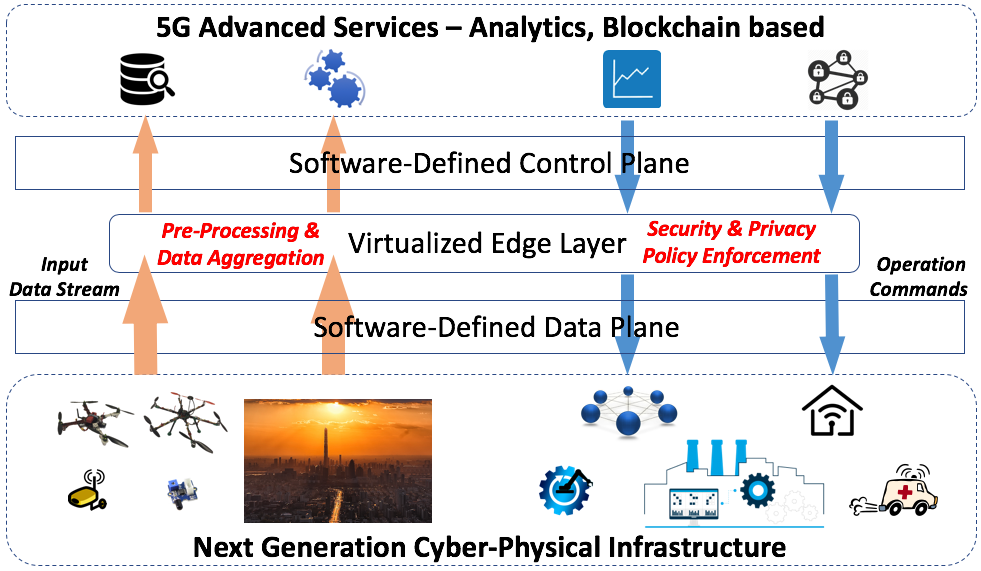}
	\caption{Edge-enabled Platform for 5G}
	\label{fig:edge}
\end{figure*}

We identify four technology advancement that can benefit 5G, including blockchain inspired distributed ledger, lightweight virtualization, software-defined Networking, and edge computing.\\

\textbf{\textit{Blockchain based Distributed Ledger}}\\

Besides upholding user privacy on the Internet, GDPR is also accelerating the development of distributed ledger technologies (i.e., blockchain based protocol design). Specifically, we are witnessing a strong demand nowadays to unify the data management across end users, companies and government. This includes providing the appropriate means to receive, track, and fulfill user requests, and to update the data as requested.

Given the high risk for enterprise in the face of steep fines, Blockchain Technology (BCT) which can store a secure historical record of transactions in a tamper-proof format will play a more visible role for many data driven applications in 5G. In this context, BCT stores information at different nodes. The past information cannot be removed and information can only be added when the nodes possess it \cite{Olnes:2017}. BCT was introduced for Bitcoin and is based on Distributed Ledger Technology (DLT), in which each participant has access to a shared ledger which is stored in many nodes \cite{Nakamoto:2008}. All transactions are stored in a ledger and all nodes have a copy of this. In turn, blocks are time stamped batches of valid transactions in which  each block includes the hash of the prior block. Creating new blocks is known as mining \cite{Narayanan:2016}. By linking the blocks a chain is formed which has resulted in the use of the name ‘blockchain’. Blockchain uses can be found in many sectors \cite{Tapscott:2016}. Especially for 5G empowered IoT and mobile financial services, we will see a merge between IoT security solutions \cite{Hafeez.2015, Hafeez.2016, Hafeez.2017, Hafeez.2017a, Hafeez.2018} and novel blockchain and cryptocurrency designs to achieve better accountability and privacy in 5G.\\

\textbf{\textit{Virtualization and Software-Defined Networking}}\\

Network Function Virtualization (NFV) is a solid technology to organize network related computation in 5G. By its design, NFV utilizes virtualization technologies to decouple physical network equipment from the functions running on them \cite{Han.2015}. This way, various virtual network functions can be implemented and deployed on one or more physical servers. In particular, the new lightweight virtualization technologies such as Docker and Unikernels \cite{Morabito.2018} will facilitate 5G to support new services of IoT domain which demands multi-tenancy, low cost, efficient resource utilization, and low power consumption.

Meanwhile, for managing 5G network traffic, Software-Defined Networking (SDN) is another powerful tool which has been successfully applied to data centers and commercial networks. In its essence, SDN decouples the data and control plane so that all the control functions can be implemented in a centralized network controller. Its design transfers the control functionality to software based entities, which eliminates the use of vendor specific back-box hardware and promotes the use commodity switches in data plane over proprietary appliances \cite{kreutz:2015}. On top of its security benefits \cite{Ding.2014}, SDN can better support multi-tenancy for large scale deployment of 5G services, such as in smart city and industrial operations, e.g., by using frameworks such as SoftOffload \cite{Ding.2015a, Ding.2015, SoftOffload}. \\

\textbf{\textit{Mobile Edge Computing}}\\

The convergence of mobile Internet and wireless systems in 5G can trigger an explosive growth in resource-hungry and computation-intensive applications, which cover a broad paradigms of IoT. These IoT systems include real-time video / audio surveillance, smart e-health, smart transportation, and Internet of Vehicles (IoV). \textit{Mobile edge computing}, by complementing various cloud resources and bringing computation closer to smart devices/objects, has been envisioned as an enabling and highly promising technology to reap the potential of IoT in 5G.

Recently, multi-access mobile edge computing (MA-MEC), which actively exploits a systematic and adaptive integration of wireless access technologies in 5G, will further enhance the access capacity between smart devices and mobile edge platforms. The design of MA-MEC is in line with the evolution towards ultra-dense deployment of small-cells (micro / pico / femto cells) in future 5G. In specific, the densely deployed 5G small cells can enhance the capacity and quality of the connections. As an example, the emerging dual-connectivity in 5G networks can enable smart objects to communicate with conventional macro-cells and offload data traffic to small cells simultaneously \cite{Ding.2011, Korhonen.2012, Ding.2012, Korhonen:2013, Ding.2013}. This enhances the access capacity of mobile edge cloud at small cells. In addition, the existing computational offloading techniques \cite{Cuervo.2010, Chun.2011, Kosta.2012}, including edge offloading \cite{Cozzolino.2017}, will further complement the needs for speeding up computation and low communication latency. \\

\textbf{\textit{Edge-enabled 5G Service Framework}}\\

To exemplify how to combine new technologies into 5G, we propose an edge-enabled platform for 5G to consolidate data management in large scale cyber-physical system deployment. As shown in Figure \ref{fig:edge}, through this platform, the 5G edge layer is expected to bring a variety of benefits, such as i) ultra-low latency between smart devices and edge cloud for real-time, interactive, and mission-critical applications, e.g., industrial operations; ii) privacy and security in local communications; and iii) fast data processing at the point of capture for IoT applications. For instance, the proposed platform will provide robust and ultra-low latency connections for smart vehicles to efficiently access the edge layer deployed on road-side units for real-time information processing. To build the edge layer, we can utilize the SDN framework \cite{SoftOffload}, IoT management tool \cite{Haus.2017a, Haus.2017b}, and Kafka framework. \footnote{https://kafka.apache.org/}

\begin{table*}[t!]
	\centering
	\caption{Comparison of 5G Pilot Initiatives}
	\label{tab:pilots}
	\begin{tabular}{|p{3cm}|C{3cm}|C{2.5cm}|C{4cm}|C{2.5cm}|}
		\hline
		{\textbf{\centering{Pilots}}}
		&\textbf{{Experiment Scale}}
		&\textbf{{Technology}}
		&\textbf{{Operation Model}}
		&\textbf{{Focus}}  \\ \hline \hline

		\rowcolor{tabcolor}
		
		Envisioned 5G Pilot &  City Scale & 5G driven & Public-Industry hybrid & Consumer and public services \\
		\hline

		Singtel 5G   &  Regional (Buona Vista area, Singapore) & 5G driven & Company driven - Ericsson (vendor) & Network services \\
		\hline
		
		Toronto Waterfront  & Regional (Port area in Toronto) & Fixed network & Company - Google (cloud service provider), and partially public sector & Infrastructure oriented \\
		\hline
		
	\end{tabular}
\end{table*}

\subsection{Open Challenges}

A combination of promising technologies like NFV and edge computing is needed to meet the demands of new applications. Nevertheless, the success of 5G still requires tackling many other challenges. \\

\textbf{\textit{Technical Challenges}}\\

For 5G network operation, security is a major concern. The role of encryption, especially the operator driven pervasive encryption \cite{RFC.8404}, has raised lots of discussions across service providers (e.g., Google, Amazon), ISPs {e.g., KPN, T-Mobile}, equipment vendors (e.g., Nokia, Ericsson) and standardization units such as IETF \footnote{https://www.ietf.org/} and ETSI \footnote{https://www.etsi.org/}.

In the context of cellular systems, the conventional network management, security operations, and performance optimization have been conducted over a large majority of data traffic flows without encryption. While unencrypted traffic could facilitate troubleshooting and management operations at all network layers, it has also made pervasive monitoring by unseen parties possible. With support from service providers (e.g., Google) and increased awareness of privacy on the Internet \cite{haus:comst:2017}, more and more traffic are encrypted in an end-to-end manner. This trend has created a challenge for 5G since existing management, operational, and security practices have depended on the availability of clear text to function. For 5G operators, it is important to investigate if critical operational practices can be met by less invasive means.

Besides conventional traffic balancing between real-time and typical web traffic \cite{Jarvinen.2013}, 5G needs also to prioritize traffic types with fine granularity. In some vertical applications the quick response is needed to avoid failure, whereas other applications response-time is less an issue. However, this traffic differentiation is correlated with the net neutrality debate whether the freedom and fairness of Internet will be affected. 

To efficiently exploit computation and storage resources at mobile edge nodes, a joint optimization of placement of computation/storage resource and cell-association with radio resource allocation are required. Such joint optimization must be self-adaptive and with minimum manual efforts. The adaptation needs to take into account time-varying environments, such as the varying wireless channel states when users move across the cells and computation/storage resource utilizations. \\

\textbf{\textit{Challenges from Regulation and Governance}}\\

As connectivity becomes a common-pool resource (CPR), there is a need for governance to manage fair usage, ensure sufficient bandwidth and scalability, enforce interoperability and give priority to certain vertical applications. For this change, regulations might interfere with the role of 5G providers in the future. Latency critical application like connected vehicle might be given priority to avoid car collisions over other applications. Also the distributed nature might demand redundant coverage of areas to avoid problems in case of mall-function. Back-up and recovery plans might be required by regulations. 

Without regulation, it is challenging to ensure proper functioning when some of the components are restricted or fail (e.g., due to market failure). For instance, the spectrum allocation is open to discussion as some spectra are already occupied by Department of Defense applications in certain countries. Failures can be disastrous for critical vertical applications areas. Edge computing architectures might be needed to be able to operate independently of the network to avoid failure of the larger system. Also security should be enforced in such a way that the whole system cannot be breached by a hack. Another aspect will be interoperability between different providers and platforms. roaming between providers should be possible to ensure proper functioning of the vertical applications which are likely to be operated by multiple 5G providers.

The General Data Protection Regulation (GDPR) represents the largest change to European Union (EU) data protection laws in decades. For 5G applications, one major criteria is on the private data collected from both end users and physical infrastructure. Privacy-by-design should be guaranteed when using the 5G applications. In addition for better integration of 5G, we also need to draw lessons from studies on standardization \cite{Ding:CCR:2014}, ambidexterity \cite{Matheus.2017},  and applying open data to smart cities \cite{Janssen.2015}.

\section{Initiative and Outlook}
\label{sec:outlook}
A main challenge for 5G will be to create a network architecture that adapts to fluctuating traffic patterns, consists of promising technologies like edge computing, software defined networking, virtualization, and combines wired and wireless elements to deal with the requirements of various vertical industries. The vertical industries yield various requirements on 5G and the actual usage might fluctuate. This requires that the architecture is dynamic, able to prioritize traffic, and can ensure that edge computing power (as envisioned in Figure 4) is available for fast and efficient processing and response. 

Reflecting on our main pursuit of this paper,
\textbf{\textit{"How do we consolidate 5G driven applications across multiple vertical industries to unveil the full potential of 5G?"}}
we believe that this answer is non-trival and the answer shall be sought from developing a comprehensive piloting testbed integrating the various technologies and in which vertical industries are involved. 

This envisioned network testbed pilot needs to integrate various technologies and be compliant to regulation and governance. 
To shed light on the 5G pilot, which will combine the efforts with Delft Green Village \footnote{https://www.thegreenvillage.org/},
we compare it against the Singtel 5G initiative \footnote{https://www.singtel.com/about-Us/news-releases/journey-to-5g-singtel-and-ericson-to-launch-singapores-first-5g-pilot-network} and Toronto Waterfront \footnote{https://sidewalktoronto.ca/}.
We summarize our observations in Table \ref{tab:pilots} in terms of experiment scale, driven technology, operation model, and project focus.

Given the challenges we outlined, this pilot project must bridge the gap between research community, industrial stakeholders, and governmental institutes. In particular from technical perspectives, the envisioned 5G pilot should allow us to:
1) experiment novel radio access technologies and their feasibility for different 5G applications; 
2) incubate novel applications by creating a trail infrastructure before entering mass market;
3) expose unforeseen limitations of network configurations; and
4) illustrate how to minimize unnecessary replacement costs through a feasible migration path, which can lead to significant deployment scale.

We must note that although our work provides an extensive sampling of existing and emerging vertical applications, this study does not attempt to cover every nuance. Further piloting can reveal new challenges and be used to understand the nature of the challenges. Besides that, the requirement analysis and technologies discussed can be applied to a broad spectrum of scenarios on top of 5G context. In addition to open challenges, our work highlights the opportunities for 5G-enabled applications from both technical and governance perspectives, which can shed light on future development for researchers, engineers and policy makers from academia, industry and government.

\section*{Acknowledgments}
 \label{sec:acknowledgments}

We thank Martin Kienzle (IBM), Inge van de Water (Gemeente Delft), and Dennis Meerburg (TU Delft) for their contributory feedback.

\bibliographystyle{ACM-Reference-Format}

\end{document}